\documentclass[twocolumn,showpacs,aps,prl,superscriptaddress]{revtex4}

\usepackage{graphicx}
\usepackage{dcolumn}
\usepackage{amsmath}
\usepackage{epsfig}

\begin{document}

\title{
  {\large 
    \bf \boldmath 
    On the Measurement of the Unitarity Triangle Angle $\gamma$
    from $B^0 \to DK^{*0}$ Decays
  }
}

\author{Tim Gershon}
\affiliation{Department of Physics, University of Warwick, Coventry CV4 7AL, United Kingdom}

\date{\today}

\begin{abstract}
The decay $B^0 \to DK^{*0}$ is well-known to provide excellent potential
for a precise measurement of the Unitarity Triangle angle $\gamma$
in future experiments.
It is noted that the sensitivity can be significantly enhanced by 
studying the amplitudes relative to those of the flavour-specific decay
$B^0 \to D_2^{*-}K^+$, which can be achieved by analyzing the 
$B^0 \to D\pi^-K^+$ Dalitz plot.
\end{abstract}

\pacs{13.25.Hw, 12.15.Hh, 11.30.Er}

\maketitle

Among the fundamental parameters of the Standard Model of particle physics,
the angle $\gamma = {\rm arg}\left(-V_{ud} V^*_{ub}/V_{cd}V^*_{cb}\right)$ of 
the Unitarity Triangle formed from elements of the Cabibbo-Kobayashi-Maskawa 
quark mixing matrix~\cite{Cabibbo:1963yz,Kobayashi:1973fv}
has a particular importance.
It is the only $CP$ violating parameter that can be measured using 
only tree-level decays,
and thus it provides an essential benchmark in any effort to understand
the baryon asymmetry of the Universe.
The precise measurement of $\gamma$ is one of the main objectives 
of planned future $B$ physics experiments 
(see, for example,~\cite{Browder:2007gg,Buchalla:2008jp,Browder:2008em}).

A method to measure $\gamma$ with negligible theoretical uncertainty was
proposed by Gronau, London and Wyler
(GLW)~\cite{Gronau:1990ra,Gronau:1991dp}.
The original method uses $B \to DK$ decays, 
with the neutral $D$ meson reconstructed in $CP$ eigenstates.
It was noted that the method can be extended to use $D$ meson decays to 
any final state that is accessible to both $D^0$ and $\overline{D}{}^0$,
and a number of potentially useful modes, 
including doubly-Cabibbo-suppressed decays such as $K^+\pi^-$~\cite{Atwood:1996ci,Atwood:2000ck},
multibody decays such as $K^0_S\pi^+\pi^-$~\cite{Giri:2003ty,Poluektov:2004mf}
and others~\cite{Grossman:2002aq,Rademacker:2006zx} have been proposed.

\begin{figure}[!htb]
  \includegraphics[width=0.49\columnwidth]{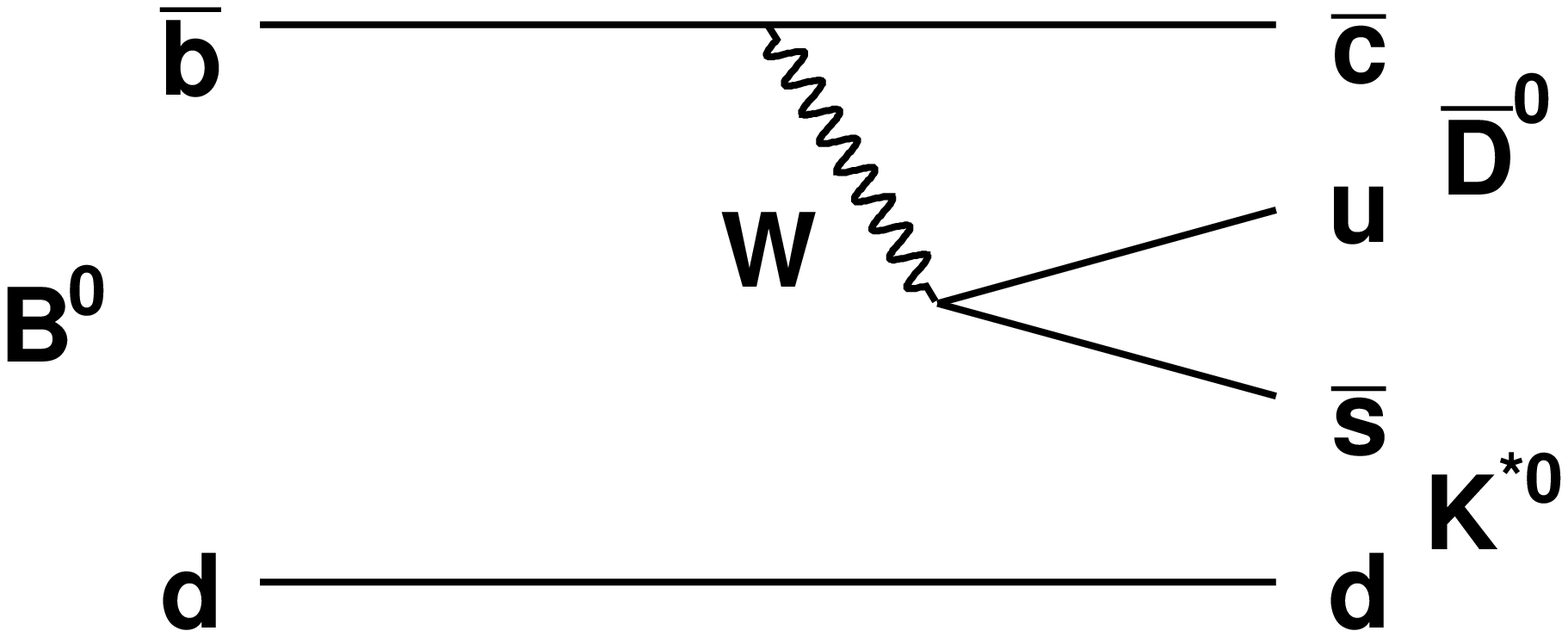}
  \includegraphics[width=0.49\columnwidth]{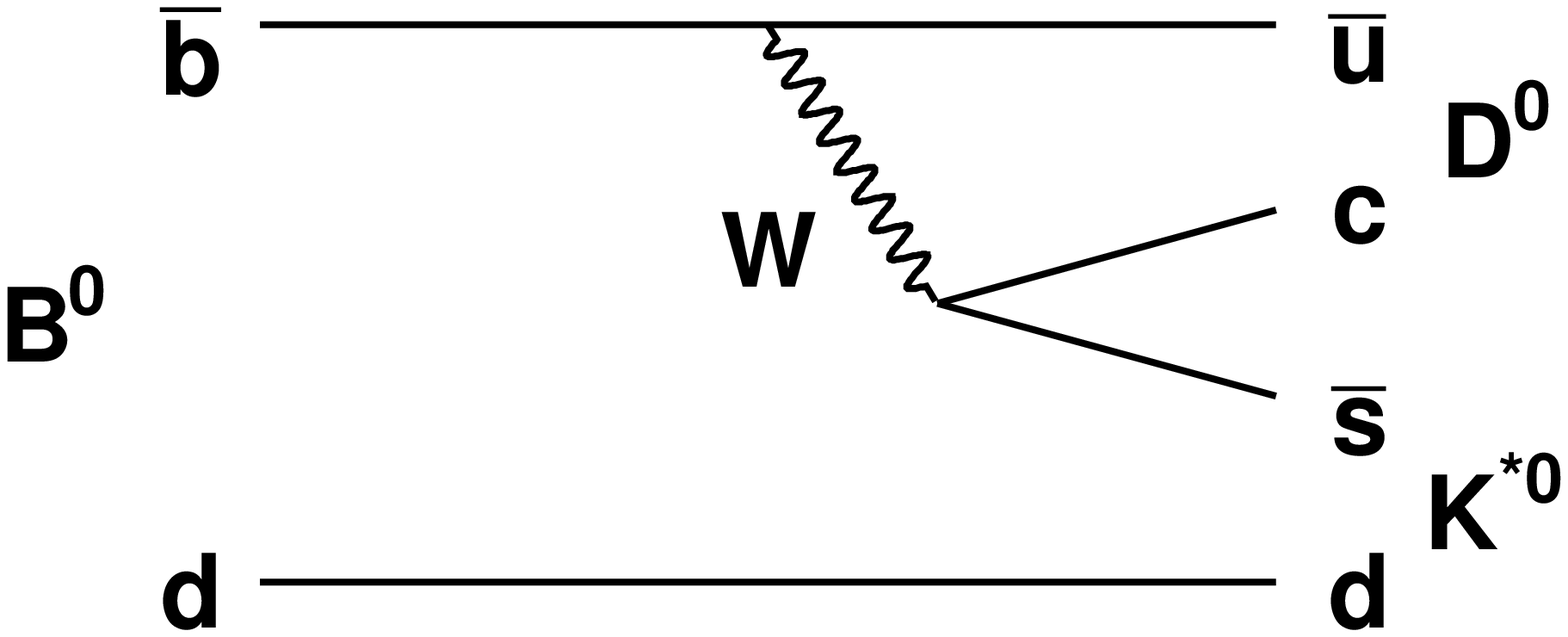}
  \caption{
    Feynman diagrams for $B^0 \to DK^{*0}$, via 
    (left) a $\bar{b} \to \bar{c} u \bar{s}$ transition
    and (right) a $\bar{b} \to c \bar{u} \bar{s}$ transition.    
  }
  \label{fig:feynman}
\end{figure}

The method can similarly be extended to other $B$ decays,
such as $B \to D^*K$ or $B \to DK^*$.
The use of neutral $B$ decays was noted as being particularly interesting
since the amplitudes involving $D^0$ and $\overline{D}{}^0$ states may
be of comparable magnitude, as shown in Fig.~\ref{fig:feynman},
potentially leading to large direct $CP$ violation~\cite{Bigi:1988ym}.
The decay $B^0 \to DK^{*0}$ is particularly advantageous since the 
charge of the kaon in the $K^{*0} \to K^+\pi^-$ decay
unambiguously tags the flavour of the decaying $B$ meson, 
obviating the need for time-dependent analysis~\cite{Dunietz:1991yd}.
This appears to be one of the most promising channels for LHCb 
to make a precise measurement of $\gamma$~\cite{Akiba:2007zz,Akiba:2008zz}.
However, the natural width of the $K^*$ meson has, until now,
been considered a hindrance to the method,
which could be handled by the introduction of 
additional hadronic parameters~\cite{Gronau:2002mu,Pruvot:2007yd,Pruvot:2007,Sordini:2008}.

\begin{figure}[!htb]
  \includegraphics[width=0.49\columnwidth]{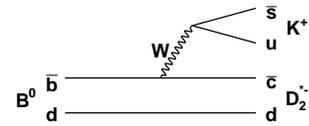}
  \caption{
    Feynman diagram for the flavour-specific $B^0 \to D_2^{*-}K^+$ decay.
  }
  \label{fig:feynman2}
\end{figure}

In this paper it is noted that the natural width of the $K^*$ meson
can be used to enhance the potential sensitivity to the $CP$ violating phase 
$\gamma$ in the analysis of $B^0 \to DK^{*0}$ decays.
By studying the $B^0 \to D\pi^-K^+$ Dalitz plots
with the neutral $D$ meson reconstructed in 
flavour-specific and $CP$ eigenstate modes,
the complex amplitudes of the $DK^{*0}$ decays can each be determined 
relative to the flavour-specific $D_2^{*-}K^+$ amplitude,
illustrated in Fig.~\ref{fig:feynman2},
allowing a direct extraction of $\gamma$ from the difference in amplitudes,
rather than from the rates.
Alternative approaches to measure $\gamma$ 
using $B \to D^{**}K$ decays~\cite{Sinha:2004ct}
or using amplitude analyses of $B \to DK\pi$ decays 
have been suggested in the literature~\cite{Atwood:1994zm,Aleksan:2002mh} 
(the time-dependent $B^0 \to D^\mp K^0_S \pi^\pm$ Dalitz plot analysis 
has recently been implemented~\cite{Aubert:2007qe}),
however the particular benefit of the  $B^0 \to D\pi^-K^+$ Dalitz plots
has not been noted until now.

Experimentally, the decay $B^0 \to \overline{D}{}^0K^{*0}$ 
has been studied by the $B$ factories,
with the world average of its branching fraction being
${\cal B}(B^0 \to \overline{D}{}^0K^{*0}) = (4.2 \pm 0.6) \times 10^{-5}$~\cite{Krokovny:2002ua,Aubert:2006qn,Barberio:2008fa}.
Initial studies of the $B^0 \to \overline{D}{}^0\pi^-K^+$ Dalitz plot 
also indicate the sizeable presence of 
the $B^0 \to D_2^{*-}K^+$ decay~\cite{Aubert:2005yt}.
Limits on the branching fraction of the $B^0 \to D^0K^{*0}$ decay 
have been set~\cite{Krokovny:2002ua,Aubert:2006qn,Barberio:2008fa},
the most restrictive limit being 
${\cal B}(B^0 \to D^0K^{*0}) < 1.1 \times 10^{-5}$ at 90\% confidence level.
First attempts to obtain constraints on $\gamma$ from $B^0 \to DK^{*0}$ decays
have been made using neutral $D$ meson decays to $K^0_S\pi^+\pi^-$~\cite{Aubert:2008yn} 
and to suppressed final states such as $K^-\pi^+$. 

To illustrate the method, consider first of all the Dalitz plot of 
the $B^0 \to \overline{D}{}^0\pi^-K^+$ decay, in which 
the $\overline{D}{}^0$ is reconstructed in the $K^+\pi^-$ final state.
Initially, this is treated as a flavour-specific decay
(hence the flavour of the $D$ meson is indicated -- 
the notation $D$ is used to indicate a neutral charm meson that is some
admixture of $D^0$ and $\overline{D}{}^0$).
The effect of the doubly-Cabibbo-suppressed $D^0 \to K^+\pi^-$ amplitude
will be considered later.
Recall that the charge of the prompt kaon in the $\overline{D}{}^0\pi^-K^+$ final state
unambiguously identifies the flavour of the decaying $B$ meson,
so that it is not necessary to consider effects due to 
$B^0$--$\overline{B}{}^0$ mixing~\cite{Gronau:2007bh}.

The $B^0 \to \overline{D}{}^0\pi^-K^+$ Dalitz plot will, of course, 
contain $\pi^-K^+$ resonances such as $K^{*0}(892)$, $K_0^{*0}(1430)$ and 
$K_2^{*0}(1430)$.  
One advantage of the Dalitz plot approach is that the hadronic parameters of
each resonance can be determined, avoiding the complications that arise due to
the use of effective hadronic parameters in the quasi-two-body $DK^*$
analysis~\cite{Gronau:2002mu,Pruvot:2007yd,Pruvot:2007,Sordini:2008}.
Furthermore, $CP$ violation effects can be studied simultaneously in all of
the contributing $\pi^-K^+$ resonances, enhancing the sensitivity to $\gamma$.
However, more importantly, the Dalitz plot will also contain significant 
contributions from $\overline{D}{}^0\pi^-$ resonances 
such as $D_0^{*-}(2400)$ and $D_2^{*-}(2460)$
(contributions from $D^{*-}(2010)K^+$ are not considered, 
since the $D^{*-}(2010)$ is too narrow to interfere with other resonances).
The crucial point is that for such resonances the flavour of the $D$ meson 
is unambiguously identified by the charge of the accompanying pion, 
independent of the $D$ decay mode.
Resonances of $\overline{D}{}^0K^+$ are not possible 
(at least, not as simple quark-antiquark mesons),
and the presence of any $D_s^{**+}$-type contributions to the Dalitz plot
would indicate the presence of amplitudes involving the $D^0$ meson.

\begin{figure}[!htb]
  \includegraphics[width=0.49\columnwidth]{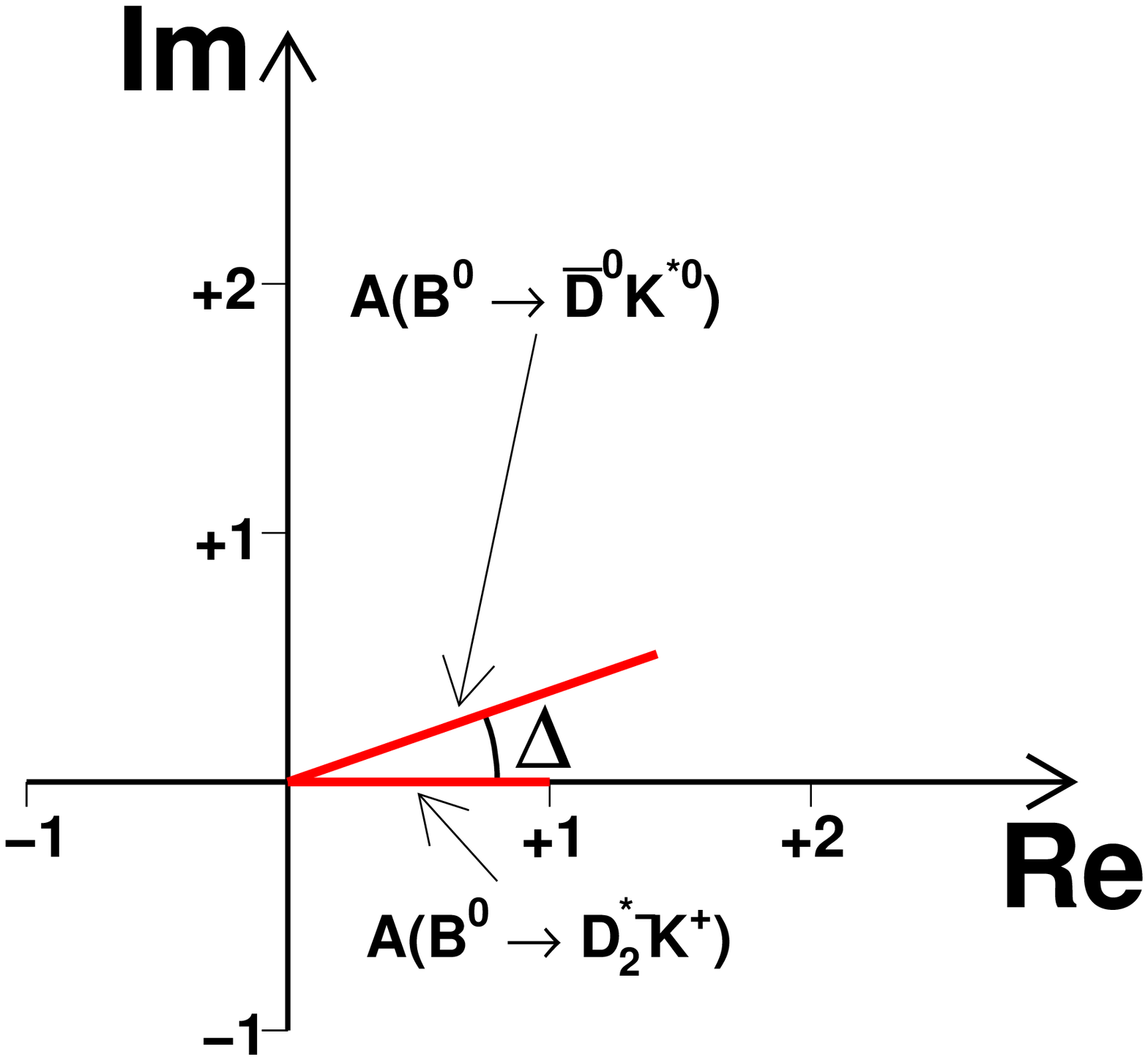}
  \includegraphics[width=0.49\columnwidth]{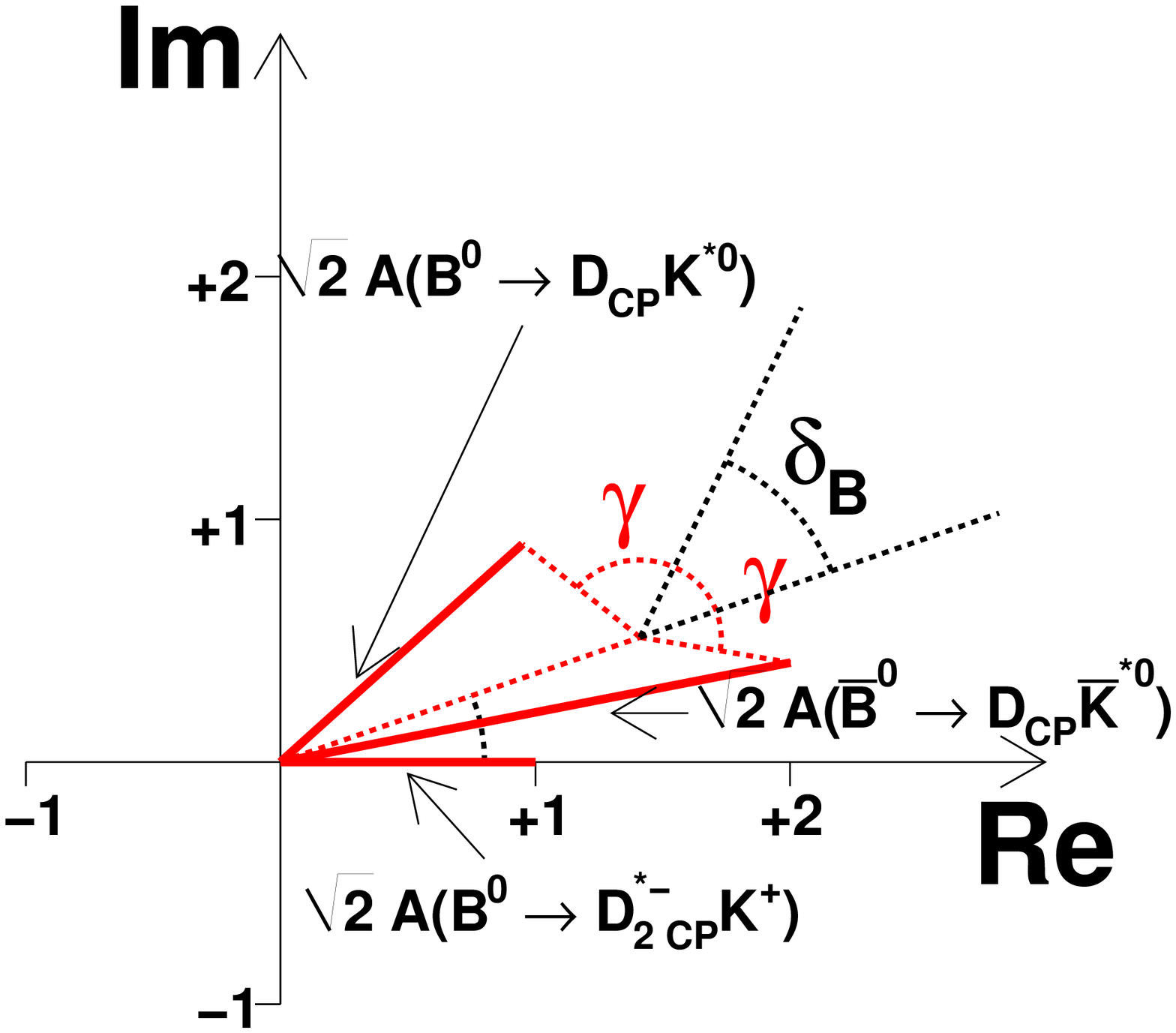}
  \caption{
    Argand diagrams illustrating the measurements of relative amplitudes 
    and phases from analysis of the Dalitz plots of 
    (left) $\overline{D}{}^0\pi^-K^+$ and (right) $D_{CP}\pi^-K^+$.
    In these illustrative examples the following values are used:
    $\left| A(B^0 \to \overline{D}{}^0K^{*0}) / A(B^0 \to D_2^{*-}K^+) \right| = 1.5$,
    $\Delta = {\rm arg}\left(A(B^0 \to \overline{D}{}^0K^{*0}) / A(B^0 \to D_2^{*-}K^+) \right) = 20^\circ$,
    $\gamma = 75^\circ$,
    $\delta_B = 45.0^\circ$ and $r_B = 0.4$.
    These are in line with expectation and current measurements,
    though $\Delta$ and $\delta_B$ are unconstrained at present.
    \label{fig1}
  }
\end{figure}

It is sufficient to consider a toy model of the Dalitz plot containing only 
$K^{*0}(892)$ and $D_2^{*-}(2460)$ resonances.
The amplitude of the $\overline{D}{}^0K^{*0}$ decay relative to that
of the $D_2^{*-}K^+$ decay can be determined, 
as illustrated in Fig.~\ref{fig1} (left), 
where the relative phase between the two amplitudes is denoted by $\Delta$.
The complex amplitudes of any other contributions to the Dalitz plot 
can be and should be determined simultaneously,
so as not to bias the extraction of the amplitudes of interest,
but this does not affect the principle of the measurement.
Since the neutral $D$ meson is flavour-specific, 
all contributions to the Dalitz plot are dominated by the 
$b \to c \overline{u} s$ tree-level transition, 
{\it ie.} all have the same weak phase.
Therefore, no direct $CP$ violation is expected and the same relation 
between amplitudes should be obtained for $B^0$ decays
and for the conjugate $\overline{B}{}^0$ decays.

Consider now the amplitudes that will be determined when a similar analysis
is applied to the $D\pi^-K^+$ Dalitz plot when the neutral $D$ meson
is reconstructed in $CP$-even eigenstates such as $D \to K^+K^-$.
(As mentioned later, $CP$-odd decays such as $D \to K_S^0\pi^0$ can also be
included in the analysis if they are experimentally accessible.)
Since the $D_2^{*-}K^+$ amplitude is flavour specific, 
the reference amplitude remains the same.
In Fig.~\ref{fig1} (right) this amplitude is denoted as 
$\sqrt{2}A(D_{2\,CP}^{*-}K^+)$ where $D_{2\,CP}^{*-}$ denotes that the neutral
$D$ meson produced in the decay of the $D_2^{*-}$ is reconstructed in a
$CP$-even eigenstate.  Neglecting trivial phase factors,
$\left| D_{CP} \right> = \frac{1}{\sqrt{2}}\left( \left| D^0 \right> + \left| \overline{D}{}^0 \right> \right)$
so that the relation $\sqrt{2}A(D_{2\,CP}^{*-}K^+) = A(D_2^{*-}K^+)$ holds.

In the absence of contributions from $D^0K^{*0}$
one would expect to find exactly the same amplitude for $DK^{*0}$
relative to that for $D_2^{*-}K^+$ as found for flavour-specific $D$ decays.
The extracted relative amplitude therefore contains information about 
the ratio of the $B^0 \to D^0K^{*0}$ and $B^0 \to \overline{D}{}^0K^{*0}$ 
amplitudes,
$r_B = \left| A(B^0 \to D^0K^{*0}) / A(B^0 \to \overline{D}{}^0K^{*0}) \right|$,
their relative strong phase difference $\delta_B$, 
and their relative weak phase difference $\gamma$.
This is illustrated in Fig.~\ref{fig1} (right), 
for both $B^0$ and $\overline{B}{}^0$ decays, 
where the sign of the weak phase difference between the amplitudes is flipped.

It is clear that the triangle constructions shown in Fig.~\ref{fig1} (right)
are exactly those regularly drawn 
to illustrate the GLW method~\cite{Gronau:1990ra,Gronau:1991dp}, 
except rotated by a constant angle $\Delta$.
To reiterate the advantage of the approach outlined here,
in the typical quasi-two-body $DK^*$ analysis,
one must reconstruct these triangles from measurements only of 
the lengths of the long sides and the base;
in this approach, one determines directly the positions 
of the apexes of the triangles.
Thus this approach provides significant additional information to constrain 
$\gamma$, as well as resolving ambiguities in the strong phase difference.
As a further elaboration of this point,
note that the rate and asymmetry measurements in the usual GLW analysis
can be translated into measurements of the parameters 
$x_{\pm} = r_B \cos(\delta_B \pm \gamma)$ conventionally used in studies of 
$B \to D^{(*)}K^{(*)}$ decays with subsequent multibody $D$ decays such as 
$D \to K^0_S\pi^+\pi^-$~\cite{Aubert:2008bd,Abe:2008wya}.
However, this is only possible if asymmetries and rates for both $CP$-even
and $CP$-odd $D$ decays have been measured, 
and furthermore no constraints on $y_{\pm} = r_B \sin(\delta_B \pm \gamma)$ 
are obtained (except indirectly from a constraint on 
$r_B^2 = x_{\pm}^2 + y_{\pm}^2$).
With Dalitz plot analysis of the $D\pi^-K^+$ Dalitz plots,
both problems are solved:
from the relation
\begin{eqnarray}
  x_+ + i y_+ & = & r_B e^{i(\delta_B + \gamma)} \nonumber \\
  & = & \frac{
    (\sqrt{2}A(D_{CP}K^{*0}))/(\sqrt{2}A(D_{2\,CP}^{*-}K^+))
  }{
    (A(\overline{D}{}^0K^{*0}))/(A(D_2^{*-}K^+))
  } - 1 \nonumber \\
  & = & \frac{\sqrt{2}A(D_{CP}K^{*0})}{A(\overline{D}{}^0K^{*0})} - 1 \, ,
  \label{eq1}
\end{eqnarray}
both $x_+$ and $y_+$ can be obtained using only $CP$-even and flavour specific
$D$ decays reconstructed in $B^0$ decays, 
with $x_- + i y_-$ similarly obtained from the conjugate $\overline{B}{}^0$
decays.
In Eq.~\ref{eq1}, the fact that both $D_{CP}K^{*0}$ and
$\overline{D}{}^0K^{*0}$ amplitudes must be determined relative to
$D_2^{*-}K^+$ is made explicit.
(If $CP$-odd $D$ decays are also used, 
to add statistics and to provide a useful experimental cross-check, 
the right-hand side of the last two relations of Eq.~\ref{eq1} 
will be multiplied by a minus sign.)
Note that all relevant normalization factors and subdecay branching fractions
are automatically taken into account since the complex amplitudes
$A(D_{CP}K^{*0})$ and $A(\overline{D}{}^0K^{*0})$ of Eq.~\ref{eq1} 
are both obtained relative to the flavour-specific $D_2^{*-}K^+$ amplitude.
Therefore this approach, which does not require reconstruction of $CP$-odd $D$
decay modes, appears highly promising for LHCb where reconstruction of states
such as $K^0_S\pi^0$ will be extremely challenging in the hadronic
environment.
Indeed, previous studies of the potential of LHCb to measure $\gamma$ from
$B^0 \to DK^{*0}$ decays~\cite{Akiba:2007zz,Akiba:2008zz} have shown a strong
dependence of the sensitivity on the unknown value of the hadronic parameter
$\delta_B$. Since the origin of this dependence is related to the absence of
information from $CP$-odd $D$ decays, it is to be expected that it will be
appreciably reduced using the Dalitz plot analysis suggested here.

The precise gain in sensitivity to $\gamma$
compared to the quasi-two-body analysis is difficult to estimate, 
since it depends on how precisely the relative phase $\Delta$ can be measured.
Dalitz plot analyses of $B^0 \to D\pi^-K^+$ have not yet been carried out,
so there is no experimental information with which to assess this issue.
However, a study of $B^0 \to \overline{D}{}^0\pi^+\pi^-$ shows that
the relative phase between $D_2^{*-}\pi^+$ and $\overline{D}{}^0\rho^0$ 
can be well-measured~\cite{Kuzmin:2006mw}.
This interference can also be exploited to obtain weak phase information,
as recently noted~\cite{Latham:2008zs}.
Furthermore, studies of $K\pi$ resonances produced in $B$ decays have revealed
a rich structure (see, for example,~\cite{Garmash:2005rv,Aubert:2008bj}).
These results provide confidence that the phase $\Delta$ can be accurately
determined, and that the Dalitz plot $B^0 \to D\pi^-K^+$ analysis 
advocated in this paper promises a substantial improvement over the 
quasi-two-body $B^0 \to DK^{*0}$ approach.
Moreover, the analysis advocated herein obtains $\gamma$ with only a
single unresolved ambiguity 
($\gamma \to \gamma + \pi$, $\delta_B \to \delta_B + \pi$), 
whereas the quasi-two-body approach suffers an eight-fold ambiguity
(note that other methods to reduce the ambiguities exist).

The discussion above has neglected the fact that 
neutral $D$ decays to $\pi^-K^+$ are not completely flavour-specific,
due to the existence of doubly-Cabibbo-suppressed $D^0 \to \pi^-K^+$ amplitudes.
The ratio of this suppressed amplitude to its Cabibbo-favoured counterpart 
has been precisely measured to be
$r_D = \left| A(D^0 \to \pi^-K^+) / A(D^0 \to \pi^+K^-) \right| = (5.8 \pm 0.2)\%$~\cite{Zhang:2006dp,Aubert:2007wf,Aaltonen:2007uc,Amsler:2008zz};
moreover the strong phase difference between these decay amplitudes 
has recently been determined by CLEOc to be
$\delta_D = (22\,^{+11}_{-12}\,^{+9}_{-11})^\circ$~\cite{Rosner:2008fq,Asner:2008ft} 
(a more precise constraint is found from a global fit including measurements
of charm mixing parameters~\cite{Barberio:2008fa}).
When the $D \to \pi^-K^+$ decay mode is used
there will therefore be a contribution from the $B^0 \to D^0K^{*0}$ amplitude 
with magnitude suppressed by $r_B \times r_D$ compared to that of 
$B^0 \to \overline{D}{}^0K^{*0}$;
the strong phase and weak phase differences will be $\delta_B + \delta_D$ 
and $\gamma$, respectively ($CP$ conservation in $D$ decay is assumed).
This could, if neglected, potentially bias the extracted value of $\gamma$.
Any bias would be small, due to the factor of $r_B \times r_D$,
but could nonetheless be significant in an era of precision measurements.

In the analysis where the suppressed $D$ decay amplitudes are neglected,
one has four observables 
(which can, for convenience, be taken to be ($x_+, y_+, x_-, y_-$)
and three unknowns ($r_B, \delta_B, \gamma$)).
Introducing suppressed amplitudes adds two more parameters ($r_D, \delta_D$)
but also adds two new observables, since one can now measure $CP$ violating 
differences between the $B^0 \to \overline{D}{}^0K^{*0}$ decay amplitude 
and its conjugate
(both measured relative to the flavour-specific $D_2^{*}K$ amplitudes).
Furthermore, external constraints on these new parameters can be used 
in the analysis.
Therefore, it is still possible to extract $\gamma$ with a precision
that should not be significantly worse than that when the suppressed 
amplitudes are neglected.
(A more precise measurement of $\delta_D$ would, however, be useful.)

One may consider whether studying the $B^0 \to D\pi^-K^+$ Dalitz plot
with the $D$ meson reconstructed in the suppressed modes will add 
additional useful information.
Although this appears promising, there will be a complication since
the flavour-specific $D_2^{*}K$ amplitude that has, until now,
been used as a reference will no longer be one of the larger contributions
to the Dalitz plot.
There could, potentially, be $D_s^{**+}$-type resonances of $DK^+$ that could 
provide an alternative flavour-specific reference,
though these would be expected to be broader than one would wish for 
such a reference amplitude.
If it were possible to use such a reference, its phase relative to 
$D_2^{*}K$ could be determined in the Dalitz plot where the $D$ meson 
is reconstructed in $CP$ eigenstates.
Thus it might be possible to use information about 
the suppressed $D$ decay modes, additional to that on the rates, 
to further improve the sensitivity to $\gamma$.

In passing, it is worthwhile to note that the method described above can 
easily be extended to $B^0 \to D^*\pi^-K^+$ decays,
where the neutral $D^*$ meson can be reconstructed in decays to either
$D\pi^0$ or $D\gamma$~\cite{Bondar:2004bi}.
However, in this case there will be an additional complication due to the
different helicity amplitudes that are possible in the $B^0 \to D^*K^{*0}$
decay~\cite{Sinha:1997zu}.
The method can also be extended to use other $D$ decays,
including multibody decays such as $D \to K^0_S\pi^+\pi^-$~\cite{Pruvot:2007yd}
or others~\cite{Rademacker:2006zx}
or single-Cabibbo-suppressed decays such as $D \to K^{*\pm}K^\mp$~\cite{Grossman:2002aq,Giri:2007et}.
Another possible extension would be to use $DK^+(n\pi)^-$ final states~\cite{Atwood:1994zm}.

Finally, it should be noted that the method discussed above does not,
unfortunately, work well when applied to charged $B$ decays.
The $K^{*+}$ produced in $B^+ \to DK^{*+}$ can decay to $K^+\pi^0$ or $K^0\pi^+$.
In the former case, $D\pi^0$ resonances do not identify the flavour of the 
$D$ meson.  
While $D_s^{**+}$-type resonances of $DK^+$ are possible, 
the amplitudes for $B^+ \to D_s^{**+} \pi^0$ decays are 
expected to be rather small (by extrapolation from published results 
on $B^+ \to D_s^{+} \pi^0$, for example~\cite{Aubert:2006xy}).
The $DK^+\pi^0$ Dalitz plot can, however, benefit from a possible alleviation 
of the suppression of the $D^0K^+\pi^0$amplitude~\cite{Aleksan:2002mh}.
In the case that the $K^{*+}$ decays to $K^0\pi^+$, neutral $DK^0$ resonances
are not possible (at least, not as simple quark-antiquark mesons),
and amplitudes for $B^+ \to D^{**+} K^0$ are expected to be negligible
(by extrapolation from limits on $B^+ \to D^{(*)+} K^0$ decays, 
for example~\cite{Aubert:2005ra}).
Thus, there is no significant flavour-specific amplitude to provide
the necessary reference point by which to obtain information about $\gamma$ 
in the Dalitz plot analysis.

In summary, it has been shown that a potentially significant improvement 
in the measurement of $\gamma$ can be achieved 
by measuring the complex amplitudes of $B^0 \to DK^{*0}$ decays
relative to that of the flavour-specific decay
$B^0 \to D_2^{*-}K^+$, which can be achieved by analyzing
$B^0 \to D\pi^-K^+$ Dalitz plots.
Compared to previously suggested techniques to measure $\gamma$ from $B^0 \to
DK^{*0}$ decays, this approach helps to resolve ambiguities, solves
problems related to interferences between various resonances while avoiding
the need for the introduction of effective hadronic parameters and provides a
potentially significant overall improvement in the sensitivity while reducing
its dependency on currently unknown parameters.
This method can be used at LHCb and other future $B$ physics experiments
to make a precise measurement of this fundamental parameter of the 
Standard Model.

I am grateful to Alex Bondar, Amarjit Soni and Jure Zupan for discussions,
and to Tom Latham, Paul Harrison, Jim Libby, Owen Long, Giovanni Marchiori, 
Achille Stocchi and Vincent Tisserand 
for discussions and comments on the manuscript.
This work is supported by the
Science and Technology Facilities Council (United Kingdom).

\end{document}